\documentclass[aps,prb,twocolumn,groupedaddress,nofootinbib]{revtex4-1}
\usepackage{amsmath,bm}
\usepackage{graphicx}
\usepackage[bottom]{footmisc}
\usepackage{float}
\usepackage{amsmath}  
\usepackage{amsfonts} 
\usepackage{graphicx} 
\usepackage{subfig}
\usepackage{bbold}
\usepackage{amsmath,amsfonts,amsthm,amssymb}
\usepackage{setspace}
\usepackage{fancyhdr}
\usepackage{extramarks}
\usepackage{bm}
\usepackage{graphicx}
\usepackage{slashed}
\usepackage[font=small, labelfont=bf,
   justification=justified,
   format=plain]{caption}
\newcommand{\tbf}[1]{{\textbf{#1}}}

\def\im{{\text{Im}\,}}
\def\Tr{{\text{Tr}\,}}

\begin{document}


\title{Density Wave Mediated Dzyaloshinskii-Moriya Interactions}

\author{Ian E. Powell, Steven Durr,  Nicholas Rombes, and Sudip Chakravarty}

\date{\today}
\begin{abstract}
We investigate the effect that density wave states have on the localized spins of a square lattice. We find that topologically nontrivial density wave states can induce stable Dzyaloshinskii-Moriya (DM) interactions among the localized spins of the lattice in the presence of an external magnetic field, and we study the resulting spin models for both antiferromagnetic and ferromagnetic backgrounds. While the density wave state itself can contribute to the the thermal Hall effect, as shown by Li \& Lee\cite{Li:2019wh} (arXiv:1905.04248v3), symmetry considerations preclude the resulting spin excitations from inducing a further thermal Hall effect. We utilize a Holstein-Primakoff (HP) substitution about the classical mean-field ground state to calculate the magnon dispersion for LSCO and find that the density wave induces a weak $d_{x^2-y^2}$ anisotropy; upon calculating the non-Abelian Berry curvature for this magnon branch we show explicitly that the magnon contribution to $\kappa_{xy}$ is zero.  Finally, we calculate corrections to the magnetic ground state energy, spin canting angles, and the spin-wave dispersion due to the topological density wave for ferromagnetic backgrounds. We find that terms \emph{linear} in the HP bosons can affect the critical behavior, a point previously overlooked in the literature.
\end{abstract}
\maketitle 

\section{Introduction}
Despite concerted efforts to illuminate the precise nature of the pseudogap phase of the cuprate high-temperature superconductors\cite{Varma:1999ca,Varma:2006jc,Yang:2006eq,Norman:2007iy}, it remains unclear which of a host of competing order parameters is responsible for the interesting behavior of this phase. One promising candidate\cite{Chakravarty:2001cl} is the $\ell=2$ spin-singlet order, the $d$-density wave (DDW), which gives rise to a $d_{x^2-y^2}$ gap and currents that alternate between adjacent plaquettes on a square lattice. The relevance of this state is certainly believable given the proximity of the pseudogap phase to the antiferromagnetic Mott insulator at low doping, which doubles the Brillouin zone in the same way and is susceptible to singlet pairing.

This density wave state of nonzero angular momentum belongs to a larger class of such states\cite{Nayak:2000kc}, and it is worth exploring other, more exotic members of this class related to the singlet DDW which maintain the key characteristics necessary for relevance to the pseudogap phase. Such states are also of some interest due to their topological properties.\cite{Hsu:2011ki} We focus on a mixed triplet-singlet DDW order, which has generated interest recently due to promising transport calculations consistent with surprising physics found in the pseudogap phase of the cuprate superconductor La${}_{2-x}$Sr${}_x$CuO${}_4$ and related compounds.\cite{Grissonnanche:2019us,Li:2019wh} Namely, for nonzero hole doping, the mixed triplet-singlet DDW state generates a nonvanishing thermal Hall conductivity $\kappa_{xy}$, and hosts hole pockets on the reduced Brillouin zone boundaries consistent with Hall coefficient measurements.\cite{Chakravarty:2008dx,Daou:2010uh}

At the mean-field level a general density wave state may be described by the Hamiltonian
\begin{equation}{\label{eq:DDWgen}}
H_{\text{DDW}} = \sum_{\tbf{k},\tbf{Q}}  c^\dagger_{\tbf{k}+\tbf{Q}}[\Phi_{\tbf{Q}}^\mu(\tbf{k})\tau^\mu]c_{\tbf{k}} + \text{h.c.},
\end{equation}
where $\tbf{Q}$ is the wave vector at which the density wave condensation occurs; $\tau^1,\tau^2,$ and $\tau^3$ are the Pauli matrices; and $\tau^0= \mathbb{1}$.
This Hamiltonian can be thought of as arising from a mean-field decomposition of nearest neighbor electron-electron interaction terms in the most general problem\cite{Laughlin:2014je,Nersesyan:1999db,Schulz:1989dx,Kampf:2003kx} in which the order parameter
\begin{equation}
\langle c^\dagger_{\tbf{k}+\tbf{Q}, \alpha} c_{\tbf{k}, \beta}\rangle = [\Phi_{\tbf{Q}}^\mu(\tbf{k})\tau^\mu]_{\alpha\beta}
\end{equation}
acquires a nonzero value for some nonzero $\tbf{Q}$.  In our work we assume that all terms which transform nontrivially under rotations and translations are captured by this mean-field decomposition.

Here we consider a specific example of Eq. \eqref{eq:DDWgen}, namely the the triplet-singlet DDW wave\cite{Hsu:2011ki} (denoted $i \sigma d_{x^2-y^2}+ d_{xy}$)
\begin{equation}{\label{eq:DDWdef}}
\begin{split}
	\Phi_{\tbf{Q}}^i(\tbf{k})&\propto iW_0N_i(\cos k_x-\cos k_y) \\
	\Phi_{\tbf{Q}}^0(\tbf{k})&\propto \Delta_0\sin k_x\sin k_y,
\end{split}
\end{equation}
where $N_i$ is a unit vector pointing along the spin quantization direction, $i=1,2,3$, and $\tbf{Q}=(\pi/a,\pi/a)$. 
In real space the Hamiltonian is written as
\begin{equation}
H_{\text{DDW}} = H_t + H_s
\end{equation}
with 
\begin{equation}
\begin{split}
	H_t &=  \frac{i W_0}{4} \sum_{i, \alpha, \beta} (-1)^{m+n}  (\tbf{N}\cdot\boldsymbol{\sigma})_{\alpha \beta} \\
	&\hspace{1cm}\times[c^{\dagger}_{i + a\hat{x}, \alpha}c_{i, \beta} - c^{\dagger}_{i + a\hat{y}, \alpha}c_{i, \beta}]
+\text{h.c.}
\end{split}
\end{equation}
and
\begin{equation}
\begin{split}
	H_{s} &= \frac{\Delta_0}{4} \sum_{i, \alpha, \beta} \delta_{\alpha, \beta} (-1)^{m+n} \\
 	&\hspace{0.5cm}\times\left[c^{\dagger}_{i + a\hat{x}+a \hat{y}, \alpha}c_{i, \beta} - c^{\dagger}_{i +a \hat{x}- a\hat{y}, \alpha}c_{i, \beta}\right]+\text{h.c.}
\end{split}
\end{equation}
The Hamiltonian $H_0 + H_{\text{DDW}}$, describes a topological Mott insulator\cite{Nayak:2000kc,Hsu:2011ki} with a quantized spin Hall conductance; it is a variant of the singlet $d$-density wave model hypothesized\cite{Chakravarty:2001cl} to explain the pseudogap phase of the cuprates.  Unlike the singlet $d$-density state, however, the mixed triplet-singlet $i \sigma d_{x^2-y^2}+ d_{xy}$-density wave state does not inherently break time reversal symmetry, yet it retains most of the signatures of the singlet $d$-density wave state.  For example, the $i \sigma d_{x^2-y^2}+ d_{xy}$-density wave wave state possesses similar hole pockets centered on the Brillouin zone diagonals which are consistent with both the measured Hall coefficient\cite{Grissonnanche:2019us} and some aspects of quantum oscillation experiments\cite{DoironLeyraud:2007bj,Sebastian:2008hp,Wang:2016ig}.  Recently, second-harmonic generation experiments have suggested that an inversion symmetry breaking is responsible for large second harmonic generation signatures in YBa${}_2$Cu${}_3$O${}_y$\cite{Zhao:2016ij} but we note that this could be due to, in principle, the quadrupole moment induced via a triplet $d$-density wave\cite{Nayak:2000kc}.  

This model was shown by Li \& Lee to produce a nonzero thermal Hall effect--however, despite considerable effort, we have not managed to exactly reproduce their plots using their parameter values and instead find a thermal Hall effect which is an order of magnitude smaller for nonzero temperatures shown here in Fig. \ref{fig:kappaXY}.  Details of the calculation are highlighted in the Appendix.  
\begin{center}
		 \includegraphics[width=0.485\textwidth]{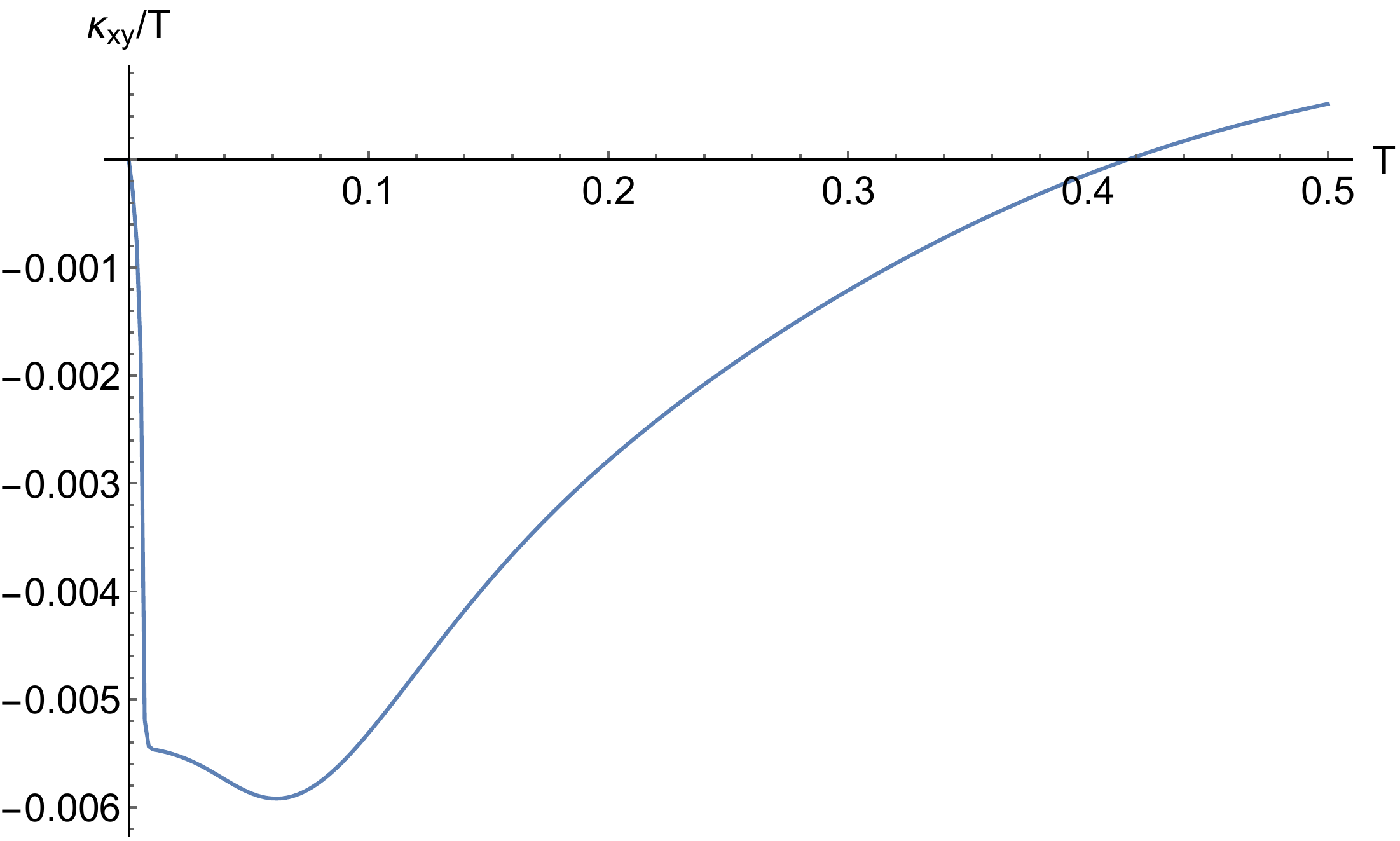}
       \captionof{figure}{
            Thermal Hall conductivity $\kappa_{xy}/T$ as a function of temperature $T$ produced by the triplet-singlet DDW state defined by Equation (\ref{eq:DDWdef}) with $\Delta_0=0.3t$, magnetic field $B=0.0075t/\mu_B$, $t' = -0.1t$, $W_0 = 2t$ at a doping of $p = 0.06$.  $\kappa_{xy}$ is listed here in units of $k_B^2/\hbar$, and $T$ in units of $t/k_B$.}
        \label{fig:kappaXY}
\end{center} 

We now ask ourselves, what effect does this density wave state have on the localized spins of the lattice? The magnitude of the experimentally-measured thermal Hall effect exceeds the maximum possible contribution from the density wave state alone by an order of magnitude; it is possible that magnetic excitations induced by the density wave state could contribute further. In our work we assume that at nonzero doping the density wave will survive in the presence of long range magnetic order\cite{Li:2019wh}.  The triplet part of the density wave order parameter induces a staggered spin current\cite{Nersesyan:1999db} on the lattice, and hence, for neighboring lattice points A and B, this intrinsic spin current implies that there exists no center of inversion at any point C on the bond connecting A and B, thereby allowing an antisymmetric exchange among the localized spins.\cite{Kikuchi:2016gy,Tatara:2019hm}  These types of (staggered) antisymmetric exchanges have been considered in the literature\cite{Kawano:2019bk}, but to our knowledge have never been considered in the context of being generated via the spin currents associated with density wave states.  

We find that the spin currents intrinsic to the triplet flavored density wave states induce a Dzyaloshinskii-Moriya (DM) interaction between the underlying neighboring spins\cite{Kikuchi:2016gy}, and we investigate the effect that this DM interaction has on antiferromagnetic and ferromagnetic spin textures, using a Holstein-Primakoff transformation expanded about the ground state. It has been previously demonstrated\cite{Samajdar:2019gk,Han:2019if,Kawano:2019bk} that certain DM interactions can lead to a thermal Hall effect. We find that the particular DM interaction induced by triplet-singlet DDW states can not contribute to $\kappa_{xy}$, which is consistent with speculations on the nature of the neutral excitation responsible for the sizable thermal Hall conductivity seen in the cuprates.\cite{Grissonnanche:2019us}

There are strong constaints and unique properties associated with the DM vectors that are generated by triplet density waves.  Because triplet density wave states break spin-rotational invariance the associated Goldstone boson excitations will destroy the two dimensional triplet density wave order at finite temperatures unless there is some external mechanism which stabilizes the triplet density wave order parameter like interlayer coupling.  However, we find that when the underlying band structure is sufficiently topologically nontrivial insofar as it hosts a nonzero spin Hall conductance, and an external magnetic field is turned on, the triplet density wave induced DM vectors are energetically stable in the absence of interlayer coupling.  Furthermore, these DM vectors are pinned to be collinear with the magnetic field, regardless of its direction, and the DM interaction will have the same symmetry as the form factor of the triplet density wave.  

In the following we derive the DM coefficients induced by triplet density waves and investigate the effects they have on the physics of the underlying spin textures of the lattice. We find that for a ferromagnetic background, the ground state remains perfectly collinear below some critical strength of the density wave; above the critical strength, the ground state acquires a nonzero canting angle, and we show that linear boson terms shift the classically-predicted threshold for nonzero canting angle.  Assuming a small DMI the dispersion of the magnons in this case develop a characteristic $d_{x^2-y^2}$ gap.  For an antiferromagnetic background we find that the density wave-induced DMI has a small effect unless there is some weak ferromagnetism present.  We compute the spin-wave spectrum approptiate to LSCO and find that for modest density wave strengths there exists a small $d_{x^2-y^2}$ anisotropy. 

\section{The Effective Magnetic Hamiltonian}{\label{sec:Hmag}}
For any type of mixed triplet-singlet density wave condensation the mean-field Hamiltonian in the absence of on-site repulsion or ferromagnetic coupling can be written in the suggestive manner
\begin{equation}
	H=\sum_{ij}c^\dagger_{i\alpha}(t_{ij}\boldsymbol{\delta}_{\alpha,\beta}+i\boldsymbol\lambda_{ij}\cdot\boldsymbol\sigma)c_{j\beta}
\end{equation}
where all singlet density wave terms are absorbed into the definition of $t_{ij}$, and $\boldsymbol{\lambda}_{ij}$ are the triplet density wave terms which couple to $\boldsymbol{\sigma}$. It can be shown\cite{Katsnelson:2010fr,Kikuchi:2016gy} that this $\boldsymbol{\lambda}_{ij}$ induces a DM interaction in the underlying spin structure (be it antiferromagnetic or ferromagnetic) whose coefficients are given by
\begin{equation}
	\tbf{D}_{ij}=\boldsymbol{\lambda}_{ij}\Tr_\sigma N_{ji},
\end{equation}
where $N_{ji}\equiv\langle c_i^\dagger c_j\rangle=-\frac{1}{\pi}\int_{-\infty}^{E_F}\im G_{ji}(E)dE$, $G_{ji}(E)$ is the Green function defined by $H$, and $E_F$ is the Fermi energy. Tracing the spin index over an expansion of $G_{ji}(E)= f(E)t_{ji}/t+g(E)\boldsymbol\lambda_{ji}\cdot\boldsymbol\sigma/t+\mathcal{O}(\lambda^2/t)$ reveals that to leading order $G_{ji}(E)$ should have the same symmetry as $t_{ji}$ under translations and rotations. 
In this work we consider a specific example of Eq. \eqref{eq:DDWgen}, written in real space as
\begin{equation}
\begin{split}
	H_{\text{DDW}} &= H_0+H_t + H_s \\
\end{split}
\end{equation}
where $H_0$ is the tight binding Hamiltonian of the underlying crystal lattice which is some union of all square planar lattices which host the triplet-singlet-DDW.
For this particular triplet-DDW case, because $\boldsymbol{\lambda}_{ij}$ only connects nearest neighbors, $t_{ji}$ is simply the tight-binding kinetic energy coefficient which we will assume to transform trivially under rotation--thus we write
\begin{equation}
	\tbf{D}_{ij}=\alpha\boldsymbol\lambda_{ij}
\end{equation}
for some constant $\alpha$.  Because we will allow the density wave strength to be a tunable parameter we will henceforth absorb $\alpha$, and all other constant numerical prefactors into the definition of $W_0$. The DM coefficients for the triplet $d_{x^2-y^2}$-density wave therefore become
\begin{equation}\label{eq:DXY}
\begin{split}
    D_{i,i\pm a\hat{x}}^{\tbf{N}}=(-1)^{i_x+i_y}W_0 \\
    D_{i,i\pm a\hat{y}}^{\tbf{N}}= -(-1)^{i_x+i_y}W_0,
\end{split}
\end{equation}
where $i = i_x + i_y$, and the superscript $\tbf{N}$ denotes that the DM vector points along the $\tbf{N}$ direction.  We stress that the method implemented here can be applied, in general, to triplet density waves in any angular momentum channel.  The direction of the DM vector is along the triplet quantization axis, and the form factor associated with the triplet density wave dictates the symmetry of the DM vector on the lattice.  To understand the order of magnitude of the DMI induced by density waves one can directly use Moriya's perturbation theory result (including the on-site repulsion $U$)\cite{Moriya:1960go}
\begin{equation}
	D \approx \frac{2 t W_0}{U},
\end{equation}
which implies that a density wave-mediated DMI is roughly on the scale of 10-100 meV for LSCO at low doping for density wave stengths $W_0 \sim t$.  

For a density wave-induced DM interaction to not be disordered by Goldstone modes at finite temperatures there must be some mechanism which externally stabilizes the triplet density wave's quantization axis, i.e. the direction of $\tbf{N}$.  Previously it was suggested that interlayer coupling was needed to stabilize the direction of $\tbf{N}$\cite{Hsu:2011ki}, however it was recently shown\cite{Li:2019wh} that the direction of $\tbf{N}$ for the triplet-singlet DDW in two dimensions can be stabilized by the bulk orbital magnetization's coupling to the magnetic field.  Explicitly, a magnetic field induces a bulk orbital magnetization, $M$, which is given by\cite{Ceresoli:2006bn}
\begin{equation}{\label{eq:bulkMag}}
M = - \sum_{\alpha = \tbf{N}\cdot\boldsymbol{\sigma} = \pm 1} \frac{e}{h c} C_{\alpha} \Delta E_{Z, \alpha},
\end{equation}
where $C_{\alpha}$ is the Chern number of the band of spin $\alpha$, $e$ is the electron charge $h$ is Planck's constant, $c$ is the speed of light, and $\Delta E_{Z,\alpha}$ is the Zeeman splitting
\begin{equation}
\Delta E_{Z, \alpha} = -\alpha \mu_{B} \text{sgn}(W_0)\tbf{N}\cdot \tbf{B},
\end{equation}
where $\mu_B$ is the Bohr magneton.  For the case of the triplet-singlet DDW the resulting energy density due to the orbital magnetization-magnetic field is\cite{Li:2019wh}
\begin{equation}
\Delta \mathcal{E}_{\text{Zeeman}} = -\frac{\mu_B B^2}{\pi c}\text{sgn}(W_0\Delta_0) (\tbf{N}\cdot \hat{\tbf{B}}),
\end{equation}
which implies that it is energetically most favorable for $W_0 \Delta_0 \tbf{N} \parallel \tbf{B}$. Thus, for $\Delta_0>0$, $\tbf{B} \neq 0$, Eq. \ref{eq:DXY} necessarily becomes
\begin{equation}
\begin{split}
	D_{i,i\pm a\hat{x}}^{\tbf{B}}&=(-1)^{i_x+i_y}|W_0| \\
    D_{i,i\pm a\hat{y}}^{\tbf{B}}&= -(-1)^{i_x+i_y}|W_0|.
\end{split}
\end{equation}
From this argument alone we see that stable density wave-induced DM interactions in two dimensions can only arise from topological density waves with nonvanishing spin Hall conductance--that is, given $\Delta E_{Z, +1} = -\Delta E_{Z, -1}$, stability is only guarenteed if $C_{+1} = - C_{-1}$.  
Furthermore, because density wave-induced DM vectors must be collinear with the magnetic field, they will transform like the magnetic field under rotations and time-reversal.  This immediately implies that the corresponding magnons in the problem will have no contribution to any thermal Hall effect because of the spin rotation and time-reversal symmetry considerations if we utilize the symmetry arguments of \cite{Samajdar:2019gk}
\begin{equation}
\begin{split}
	\kappa_{xy}[J,\tbf{D}_{ij}, \tbf{B}] &= \kappa_{xy}[J,R_{\boldsymbol\phi}\hat{\tbf{B}}D_{ij}, R_{\boldsymbol\phi}\tbf{B}] \\
	\kappa_{xy}[J,\tbf{D}_{ij}, \tbf{B}] &= - \kappa_{xy}[J,-\hat{\tbf{B}}D_{ij}, -\tbf{B}],
\end{split}
\end{equation}
where $R_{\boldsymbol\phi}$ is the vector representation of spin rotation by some angle $\phi$ about the axis defined by $\hat{\boldsymbol{\phi}}$.   Rotating the system about an angle $\pi$ about an axis perpendicular to $\hat{\tbf{B}}$ maps $R_{\boldsymbol\phi}\hat{\tbf{B}}$ to $-\hat{\tbf{B}}$ and hence $\kappa_{xy} = -\kappa_{xy} = 0.$ 
The bulk magnetization (Eq. \ref{eq:bulkMag}) would, in principle, produce a small ferromagnetic-like signal detectable in polar Kerr measurements so long as the external magnetic field is not exactly zero for weak disorder at small enough temperatures.  More detailed calculations involving interlayer coupling, the inclusion of magnetic impurities, and nonzero temperatures should be considered in future work to quantitatively compare this triplet-singlet DDW bulk magnetization signal to the polar Kerr rotation data previously gathered\cite{Xia:2008cc}.  Interesting questions to ask are how the Goldstone modes would disorder the DM vectors in the absence of an external magnetic field, and how the DM vectors behave for density wave states with vanishing spin Hall conductances. 

We now study the effect of this dynamically generated DM interaction on the isotropic two dimensional Heisenberg ferromagnet and antiferromagnet. Namely, we consider
\begin{equation}\label{hamm}
H = J \sum_{i, j} \tbf{S}_i \cdot \tbf{S}_j +\sum_{i, j} \tbf{D}_{ij} \cdot (\tbf{S}_i \times \tbf{S}_j) - \tbf{B} \cdot \sum_i \tbf{S}_i,
\end{equation}
where $J$ is the spin exchange, and the DM interaction includes the contribution from the density wave.

\section{The Antiferromagnetic Background}
The triplet-singlet density wave induced DMI will typically have little effect on perfectly antiferromagnetic backgrounds in the linear spin wave approximation.  This is because, upon turning on a magnetic field to stabilize the DMI, the localized spins will flop perpendicular to the magnetic field direction.  Terms which couple to $W_0$ in this case are proportional to higher order terms in the HP bosons.  When this happens only very large density wave strengths will cause distortions in the magnetic ordering.  On the other hand, if there exists some small ferromagnetic component associated with the otherwise antiferromagnetic ordering, the density wave-induced DMI appears in terms quadratic in the HP bosons and thus will affect the magnon dispersion.  This is the case for LSCO which we will consider in the following.

Taking $\tbf{B} = B \hat{z}$ the DM matrix for LSCO can be written as $\tbf{D}_i=(-1)^{i_x+i_y}\tbf{D}$, where
\begin{equation}
\tbf{D} = \left(\begin{matrix}
	\sqrt{2}D\cos\theta_d &\sqrt{2}D\sin\theta_d & W_0\\
	-\sqrt{2}D\sin\theta_d &  -\sqrt{2}D\cos\theta_d & - W_0\\
	0&0&0
	\end{matrix}\right).
\end{equation}
The $x$ and $y$ spin direction entries are due to the buckling of the oxygen atoms out of the copper oxide plane and induce a weak net ferromagnetic moment out of the copper oxide plane.\cite{Cheong:1989kf,Coffey:1991hn,Thio:1994fk}  We find the mean-field ground state by summing the classical energy over the four sublattices and then numerically minimizing this energy with respect to the four sets of spherical angles which characterize the classical spin directions.  Here we parameterize the spins as 
\begin{equation}
\langle \tbf{n}_i \rangle = (\text{cos}\phi_i, \text{sin}\phi_i, n_z)
\end{equation}
where $n_z$ is the weak ferromagnetic canting due to $D$.  For density wave strengths on the order of $t$ the ground state is unchanged--characterized by antiferromagnetic spin flopping in the plane perpendicular to the magnetic field with a weakly ferromagnetic component pointing along $B$ induced by $D$.  Closely following the work of Han, Park, and Lee\cite{Han:2019if}, we choose the form of the classical ground state as
\begin{equation}
\langle \bar{\tbf n}_{i}\rangle =  - \frac{(-1)^{i_x+i_y}}{\sqrt{2}}\cos{\theta_c}(\hat{x}+\hat{y}) + \text{sin}\theta_c \hat{z}.
\end{equation}
We expand our spin operators about this mean-field ground state\cite{Haraldsen:2009gi}
\begin{equation}
\tbf{S}_i = a_i \langle \tbf{n}_i \rangle + \tbf{t}_i
\end{equation}
so that we can perform the appropriate Holstein-Primakoff (HP) substitution.  The amplitudinal reduction along the mean-field state given as
\begin{equation}
a_i = S - b_i^\dagger b_i,
\end{equation}
and, to leading order in boson density operators, the transverse fluctuation operator $\tbf{t}_i$ is given by
\begin{equation}
\tbf{t}_i = t_i^{x'} \hat{x_i'} + t_i^{y'} \hat{y_i'},
\end{equation}
with
\begin{equation}
\begin{split}
	t_i^{x'} &= \sqrt{\frac{S}{2}}(b^\dagger_i + b_i) \\
	t_i^{y'} &= i\sqrt{\frac{S}{2}}(b^\dagger_i - b_i)
\end{split}
\end{equation}
where the primed coordinates are defined such that $\hat{x_i'}\times\hat{y_i'} =  \langle \tbf{n}_i \rangle$.  Upon substitution of these operators into Eq. (16) the terms quadratic in boson creation and annihilation operators in the Hamiltonian can be written in real space as (redefining the couplings to absorb $S$)
\begin{equation}
\begin{split}
H = (4 J' + B \text{sin}(\theta_c)) \sum_i b^\dagger_i b_i -  \sum_{\langle i, j \rangle} J t_i^{x'} t_j^{x'} +J' t_i^{y'} t_j^{y'}\\
&\hspace{-7.5cm}+ \sum_{i} D'[t_i^{x'} t_{i+x}^{y'}+t_i^{y'} t_{i+x}^{x'} - t_i^{x'} t_{i+y}^{y'}+t_i^{y'} t_{i+y}^{x'}]\\
&\hspace{-7cm}+ i \sum{i} W_0 (-1)^{i_x + i_y}[b_i b^\dagger_j - b^\dagger_i b_j],
\end{split}
\end{equation}
where  $J' = J/\text{cos}(2 \theta_c)$ and $D' = \text{cos}(\theta_c) D (\cos{\theta_d}+\sin{\theta_d})$.  We take the Fourier transformation and write our Hamiltonian in Nambu form as
\begin{equation}
    H=  \sum_k \frac{1}{2}\psi_k^\dagger \mathcal{H}_k \psi_k
\end{equation}
where $\psi^\dagger_k = (b^\dagger_k, b^\dagger_{k+Q}, b_{-k}, b_{-k+Q})$ and
\begin{equation}
	\mathcal{H}_k = \left(\begin{matrix}
	A_k &  2 i W_k \text{sin}\theta_c &  B_k &0\\
	0 &  -A_k & 0& - B_k \\
	0&0&A_k&  -2 i W_k \text{sin}\theta_c\\
	0&0&0&-A_k\\
	\end{matrix}\right) + \text{h.c.},
\end{equation}
where 
\begin{equation}
\begin{split}
	A_k &= 2 J' + \frac{B}{2} \text{sin}\theta_c + \frac{1}{2}(J' - J)(\text{cos}k_x + \text{cos}k_y) \\
	B_k &= -2 i D' (\cos k_x-\cos k_y), \\
	&\hspace{1cm}- (J' + J)(\text{cos}k_x + \text{cos}k_y), \\
	W_k &= W_0 (\cos k_x - \cos k_y).
\end{split}
\end{equation}
The spectrum is given by the absolute value of the eigenvalues of the dynamic matrix $K = (\sigma_3 \otimes \mathbb{I}_2) \mathcal{H}_k$\cite{Colpa:1978ku}-- these eigenvalues correspond to what are called particle and hole bands for the positive and negative eigenvalues respectively. We plot the dispersion in Fig. \ref{fig:disp5} for a set of parameter values.   Taking into account both the particle and hole bands the dispersion consists of one four-fold degenerate magnon branch very similar to that which is obtained when $W_0 = 0$; because $W_0$ couples to the weak ferromagnetic moment its effects on the spectrum are small if the density wave strength is not very large.  Namely, tuning $W_0$ from zero induces a weak anisotropy in the the magnon dispersion along the $k_x$ and $k_y$ directions.  The dispersion along $k_d = k_x = k_y$ is unchanged by increasing $W_0$ whereas the energy is increased along the $\tbf{k} = (k_x, 0)$, $\tbf{k} = (0, k_y)$ directions.  This anisotropy is demonstrated in Fig. \ref{fig:dispdiff}, where the largest deviation from the unperturbed magnon dispersion is roughly $0.4$ meV when $k = (0, \pi)$ and $k = (\pi, 0)$.  
\begin{center}
		 \includegraphics[width=0.485\textwidth]{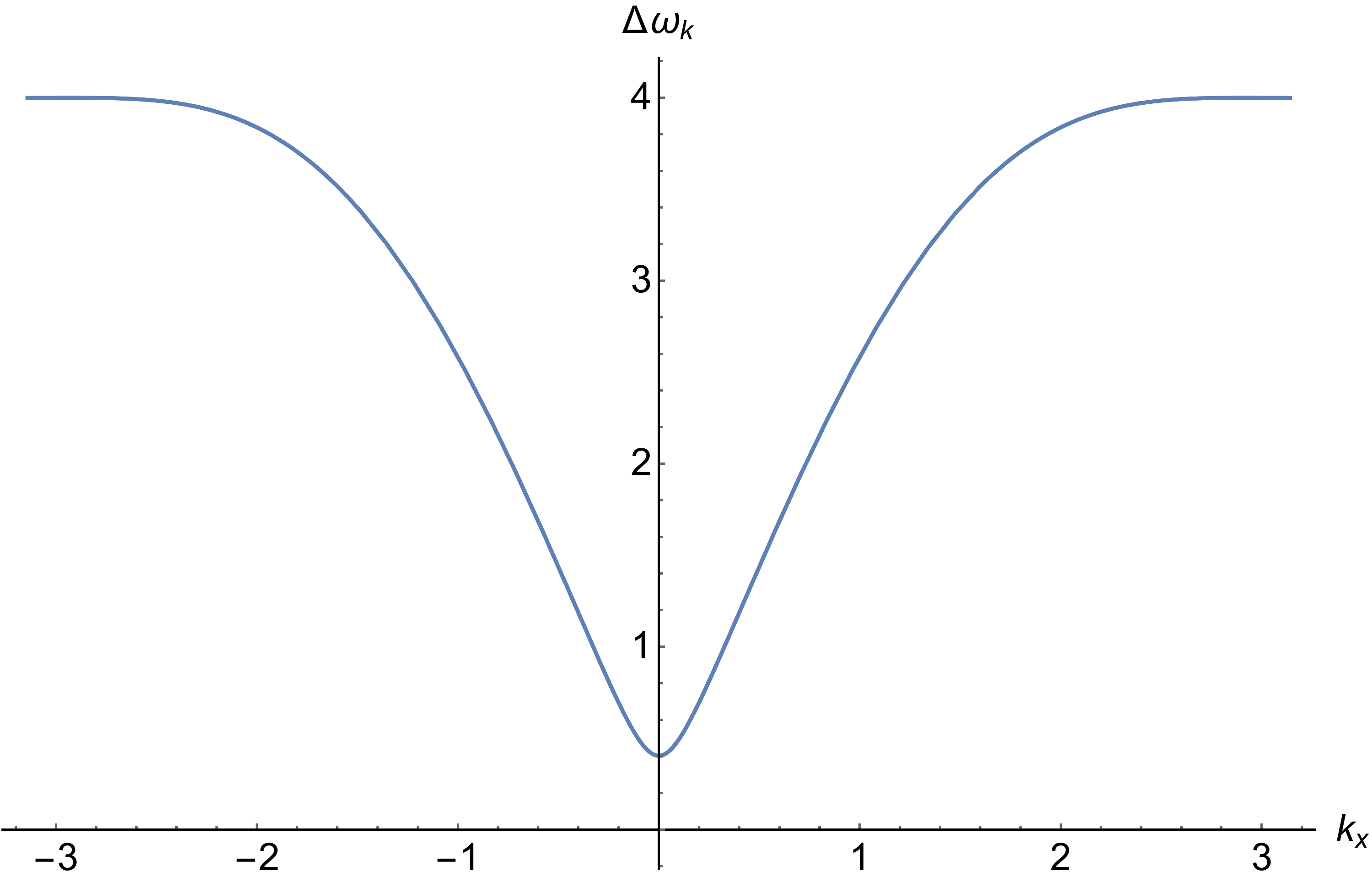}
        \captionof{figure}{
                Magnon dispersion $\omega_{k}$ for $k_y = 0$ in units of $J$ for $W_0$ = 0.3$J$, $D$ = 0.1$J$, $\theta_d$ = 0.05, $B = 0.05 J$.  The spectrum remains antiferromagnetic with small corrections increasing with the value of $k_x$.
        }
        \label{fig:disp5}
\end{center}
\begin{center}
		 \includegraphics[width=0.485\textwidth]{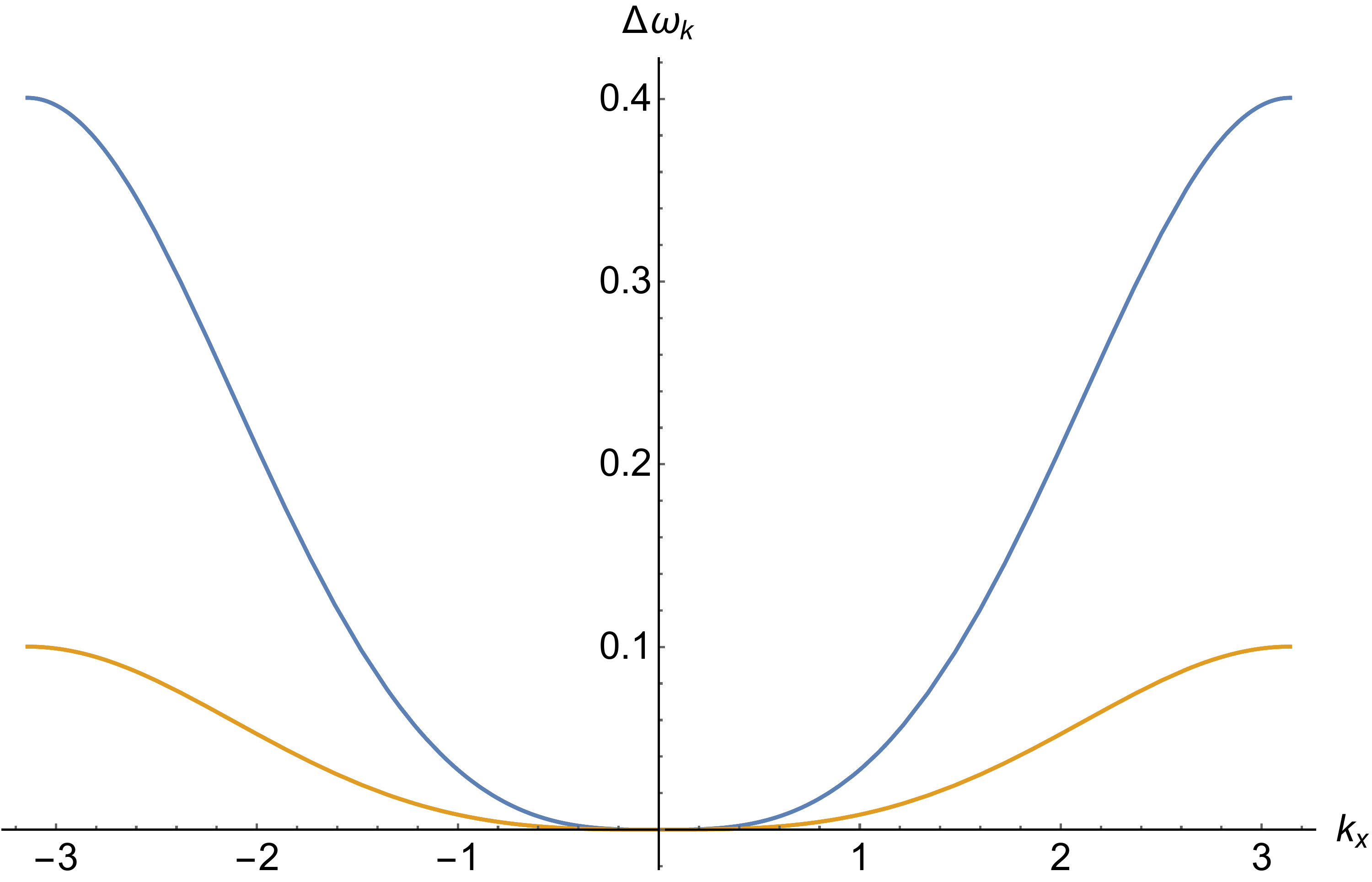}
        \captionof{figure}{
                \label{fig:dispdiff}
                The difference in the magnon dispersions listed in meV with $k_y = 0$, $D$ = 12 meV, $\theta_d$ = 0.05, and $B$ = 6 meV.  The orange, and blue curves correspond to the difference between the $W_0$ = 100 meV and $W_0$ = 0 dispersions and the difference between the $W_0$ = 50 meV and $W_0$ = 0 dispersions respectively.
        }
\end{center} 
\begin{center}
		 \includegraphics[width=0.485\textwidth]{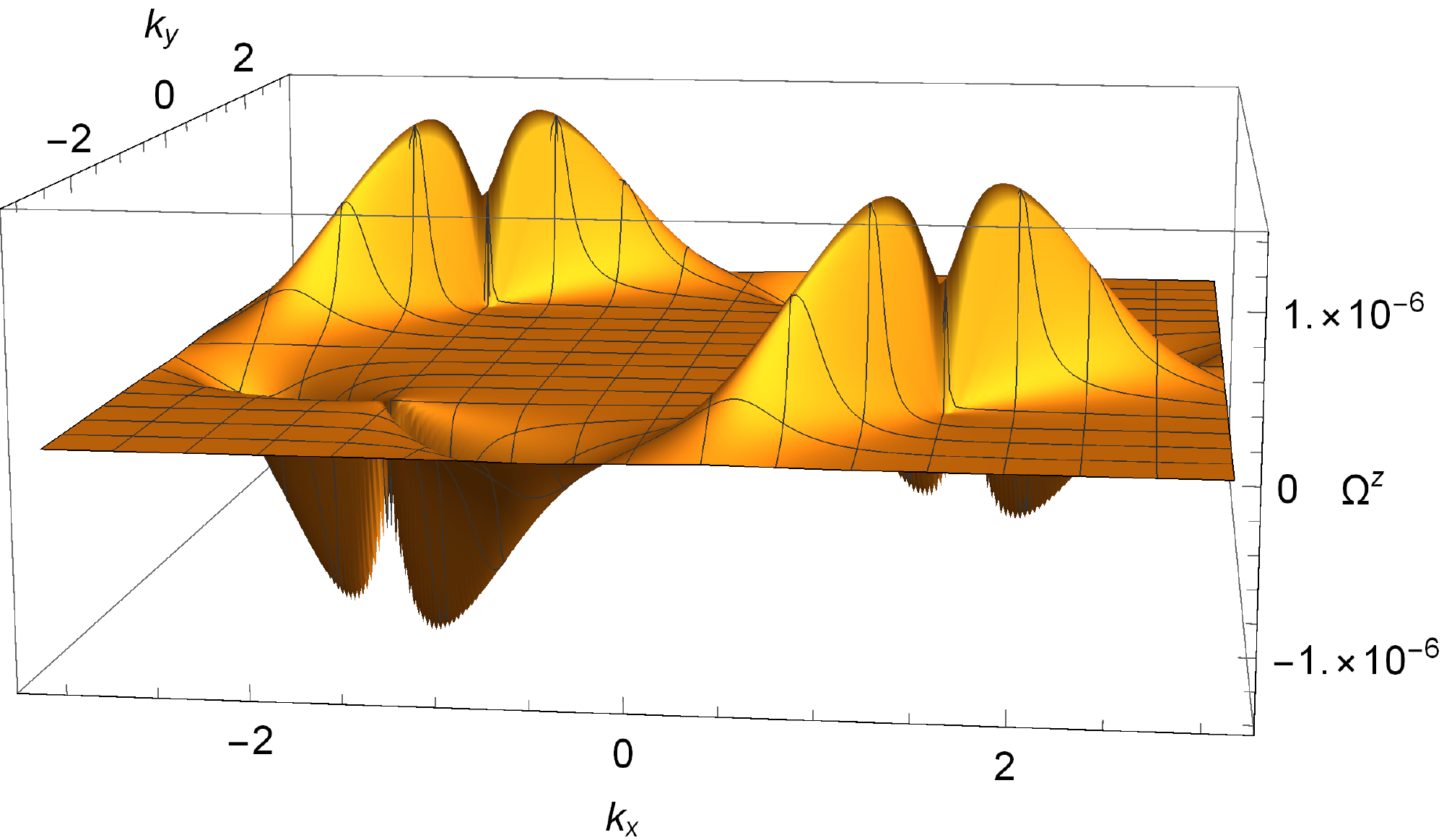}
        \captionof{figure}{
                \label{fig:NABC}
               	 Non-Abelian Berry curvature calculated on a 200$\times$200 lattice with $W_0$ = 0.3$J$, $D$ = 0.1$J$, $\theta_d$ = 0.05, $B = 0.05 J$.  
        }
        \label{fig:disp6}
\end{center} 
The thermal Hall conductivity is given by\cite{Matsumoto:2014ex}
\begin{equation}
\begin{split}
	\frac{\kappa_{xy}}{T} &= -\sum_n\frac{k_B^2}{\hbar}\int \frac{d\tbf{k}}{(2\pi)^2} \\
	&\hspace{1cm}\left[c_2(n_B(\omega_{n, \tbf{k}}))-\frac{\pi^2}{3}\right]\Omega_n^z(\tbf{k})\label{eq:th}
\end{split}
\end{equation}
where 
\begin{equation}
c_2(x) = \int_0^x ds \left(\text{ln}\left[\frac{1+s}{s}\right]\right)^2,
\end{equation}
$n_B$ is the Bose distribution function, $\Omega_n^z$ is the Berry curvature of the $n$'th band, and the sum is taken over the particle bands. 
Because the particle band in this model is twofold degenerate we calculate its non-Abelian Berry curvature using the discretized link method.\cite{Fukui:2005er}
We define the Berry curvature as
\begin{equation}
\Omega^z(\tbf{k}) = - i \ln[\tilde{U}_1(\tbf{k})\tilde{U}_2(\tbf{k}+\tbf{e}_1)\tilde{U}_1(\tbf{k}+\tbf{e}_2)^{-1}\tilde{U}_2(\tbf{k})^{-1}] 
\end{equation}
where the vectors $\tbf{e}_1 = 2\pi(1,0)/N$, $\tbf{e}_2 = 2\pi(0,1)/N$, and $N^2$ is the total number of lattice sites.  The link variables are defined as
\begin{equation}
 \tilde{U}_\gamma(\tbf{k}) = \frac{\text{det}U_\gamma(\tbf{k})}{|\text{det}U_\gamma(\tbf{k})|}
\end{equation}
where the matrix entries of $U_\gamma (\tbf{k})$ are the eigenstate overlap elements in the degenerate subspace, which for the magnon case take the form\cite{Shindou:2013cp}
\begin{equation}
U_\gamma (\tbf{k}) = \left(\begin{matrix}
	\langle \psi_1(\tbf{k})|\tilde{\psi}_1(\tbf{k}+\tbf{e}_\gamma)\rangle &\langle \psi_1(\tbf{k})| \tilde{\psi}_2(\tbf{k}+\tbf{e}_\gamma)\rangle \\
	\langle \psi_2(\tbf{k})|\tilde{\psi}_1(\tbf{k}+\tbf{e}_\gamma)\rangle & \langle \psi_2(\tbf{k})|\tilde{\psi}_2(\tbf{k}+\tbf{e}_\gamma)\rangle \\
	\end{matrix}\right).
\end{equation}
Here the magnon eigenstates $|\psi_i(\tbf{k})\rangle$ are the normalized eigenvectors of $K$ that correspond to the positive energy eigenvalues--i.e. the particle bands, and $|\tilde{\psi}_i\rangle = (\sigma_3 \otimes \mathbb{I}_2)|\psi_i\rangle$.
We plot the non-Abelian Berry curvature calculated on an $200\times200$ lattice in Fig. \ref{fig:NABC} for a characteristic set of parameter values.  Assuming that Eq. \eqref{eq:th} can be generalized trivially for the case of non-Abelian Berry curvature it can be seen immediately that $\kappa_{xy}$ = 0 because our numerically calculated $\Omega^z(\tbf{k})$ obeys $\Omega^z(k_x, k_y)$ = -$\Omega^z(-k_x, k_y)$ = -$\Omega^z(k_x, -k_y)$ whereas $\omega(k_x, k_y)$ = $\omega(-k_x, k_y)$ = $\omega(k_x, -k_y)$--thereby causing the integral to vanish.  

\section{The Ferromagnetic Background}
It has previously been shown\cite{Kampf:2003kx} via a one-loop renormalization group analysis of the extended $U$-$V$-$J$ model that triplet $d_{x^2-y^2}$-density wave condensation is energetically favorable for a range of interaction strengths given $J/U<0$.  Furthermore, it was theoretically predicted\cite{Kopp:2007fn} and recently experimentally confirmed\cite{Sarkar:2020ce} that the highly overdoped cuprates show ferromagnetic ordering in the CuO$_2$ planes.  Due to these reasons we investigate the mixed triplet-singlet density wave DMI effects on a two-dimensional Heisenberg ferromagnet.  Taking $\tbf{B}=B \hat{z}$ the symmetric exchange term favors mean-field states of the form
\begin{equation}
\langle \bar{\tbf S}_{i}\rangle =  n_z \hat{z},
\end{equation}
and the antisymmetric exchange favors mean-field states of the form
\begin{equation}
\begin{split}
	\langle \tilde{\tbf S}_{i}\rangle &=\xi_x(\tbf r_{i_x,i_y}) \hat{x} + \xi_y(\tbf r_{i_x,i_y}) \hat{y}, \\ 
\xi_x(\tbf r_{i}) &= \xi_0 \frac{[(-1)^{i_x}+(-1)^{i_y}]}{2} \\ 
 \xi_y(\tbf r_{i}) &= \xi_0 \frac{[(-1)^{i_x}-(-1)^{i_y}]}{2}.
\end{split}
\end{equation}
Thus, the mean-field state which occurs in the presence of both types of exchange is
\begin{equation}
 \langle \tbf{n}_{i} \rangle =  \langle \bar{\tbf{S}}_{i_x,i_y} \rangle + \langle \tilde{\tbf{S}}_{i_x,i_y} \rangle.
\end{equation}
We have checked that this is the true mean-field ground state by summing the classical energy over the four sublattices and then numerically minimizing the energy with respect to the four sets of spherical angles which characterize the classical spin directions.
The mean-field energy per site in this case is (restoring $S$)
\begin{equation}
\begin{split}
	\frac{E_0}{N} &= -|J|  z S^2 \text{cos}^2(\theta)/2 - 2  S^2 |W_0| \text{sin}^2(\theta) \\
	&\hspace{2cm}-  B S \text{cos}(\theta),
\end{split}
\end{equation}
where $N$ is the number of lattice sites, $z$ = 4 (6) in two (three) dimensions, and $\theta$ is defined as the angle between $\langle \bar{\tbf{S}}_{i_x,i_y} \rangle$ and $\langle\tbf{n}_i\rangle$.  For the square lattice case the ground state is minimized about $\theta = 0$ for all $2 W_0 < z|J|/2+B/S$, whereas the ground state is minimized at $\theta = \text{cos}^{-1}[B/(4S(W_0-J))]$ for $2 W_0 > z|J|/2+B/S$. 

Following the procedure highlighted in the previous section we expand the operators about the mean-field ground state and substitute them into Eq. \eqref{hamm} to yield the real space Hamiltonian
\begin{equation}
H = E_0  + H_0 + H',
\end{equation}
where the classical mean field energy $E_0$ is defined in Eq. (20), $H_0$ is
\begin{equation}
\begin{split}
	H_0 &=\sum_i \mu b^\dagger_i b_i + \sum_{\langle i, j \rangle}[\bar{Z}_\theta g(j) b^\dagger_i b_j + \tilde{Z}_\theta g(j) b^\dagger_i b^\dagger_j \\
	&\hspace{1cm}+ i J S \text{cos}(\theta) (-1)^{i_x+i_y} b_i b^\dagger_j +\text{h.c.}],
\end{split}
\end{equation}
with $g(j) = +1$ for $j = i+\hat{x}$, and $g(j) = -1$ for $j = i+\hat{y}$, and the coefficients are defined as
\begin{equation}
\begin{split}
	\bar{Z}_\theta &\equiv \frac{JS}{2} \text{sin}^2(\theta) + \frac{W_0 S}{2}(\text{cos}^2(\theta)+1) \\
	\tilde{Z}_\theta &\equiv \frac{JS}{2} \text{sin}^2(\theta) + \frac{W_0 S}{2}(\text{cos}^2(\theta)-1) \\
	\mu &\equiv 4J S \text{cos}^2(\theta) + 4 W_0 S \text{sin}^2(\theta) + B \text{cos}(\theta),
\end{split}
\end{equation}
and $H'$ is
\begin{equation}
H' = \sum_{i} (-1)^{i_x}  A_\theta(b_i^\dagger+b_i),
\end{equation}
where 
\begin{equation}
A_\theta = \text{sin}(\theta)\left[\sqrt{\frac{S}{2}}B + \left(J(2S)^{3/2} -  W_0(2S)^{3/2}\right)\text{cos}(\theta)\right].
\end{equation}
Terms linear in boson creation and annihilation operators imply spin-wave creation and annihilation from the ground state.  Thus, assuming that the system is in its ground state, it is typically argued in the literature that this coefficient $A_\theta$ must vanish at each point $i$ on the lattice; in the following we show that allowing small $A_\theta$ has a nontrivial effect on the critical behavior of the system.  

There exist two unique solutions for vanishing $A_\theta$: the perfectly ferromagnetic case of $\theta = 0$, and
\begin{equation}
|\theta| = \text{cos}^{-1}\left[\frac{B}{4S(W_0-J)}\right]
\end{equation}
which is the aforementioned magnetic ground state canting angle that occurs at the classical mean-field level when $2W_0 >2J+B/S$. 
Anticipating quantum corrections to the ground state canting angle we opt to include the effects of $A_{\theta} \neq 0$--however, to maintain consistency with the Holstein-Primakoff substitution about the mean-field ground state it is understood that $A_\theta$ is necessarily either small or exactly zero, i.e. we are expanding sufficiently close to the classical mean-field theory's predicted relationship between the parameters.  Thus, instead of taking $A_\theta = 0$, we eliminate terms linear in bosonic creation and annihilation operators by performing the canonical transformation
\begin{equation}
\begin{split}
	b_i &= \tilde{b_i} - (-1)^{i_x}x \\
	b_i^\dagger &= \tilde{b}^\dagger_i - (-1)^{i_x}x,
\end{split}
\end{equation}
where $x_i$ is the C-number
\begin{equation}
x= \frac{-A_\theta }{4\bar{Z}_\theta + 4\tilde{Z}_\theta-\mu}.
\end{equation}
Note that this transformation is well defined when the denominator $4\bar{Z}_\theta + 4\tilde{Z}_\theta-\mu \neq  0$--this is indeed the case when we investigate the physics in close proximity to the mean-field behavior.
The Hamiltonian then becomes
\begin{equation}
H = E_0 - N x A_\theta + H_0' = E_0' + H_0',
\end{equation}
where $H_0'$ is identical to the Hamiltonian written in Eq. (27) but in terms of the transformed bosonic operators $\tilde{b}$.  Upon Fourier transformation the total Hamiltonian can be written in terms of the Nambu basis as
\begin{equation}
    H=  E_0''-N x A_\theta + \sum_k \frac{1}{2}\psi_k^\dagger \mathcal{H}_k \psi_k
\end{equation}
where $\psi^\dagger_k = (\tilde{b}^\dagger_k, \tilde{b}^\dagger_{k+Q},\tilde{b}_{-k}, \tilde{b}_{-k+Q})$,
\begin{widetext}
\begin{equation}
	\mathcal{H}_k = \left(\begin{matrix}
	\bar{Z}_{\theta, k} +\mu/2 &   i J_k \text{cos}(\theta)&2 \tilde{Z}_{\theta, k} &0\\
	0 &  -\bar{Z}_{\theta, k} +\mu/2 & 0&-2 \tilde{Z}_{\theta, k} \\
	0&0&\bar{Z}_{\theta, k} +\mu/2& - i J_k \text{cos}(\theta) \\
	0&0&0&-\bar{Z}_{\theta, k} +\mu/2\\
	\end{matrix}\right) + \text{h.c.},
\end{equation}
\end{widetext}
\begin{equation}
\begin{split}
	\frac{E_0''}{N} &= -2 S(S+1)[ J \text{cos}^2(\theta) +|W_0| \text{sin}^2(\theta)] \\
	&\hspace{2cm}-  B (S+1/2) \text{cos}(\theta),
\end{split}
\end{equation}
and $\bar{Z}_{\theta, k} = \bar{Z}_{\theta}[\text{cos}k_x - \text{cos}k_y]/2$, $\tilde{Z}_{\theta, k} = \tilde{Z}_{\theta}[\text{cos}k_x - \text{cos}k_y]$, $J_k = J S [\text{cos}k_x + \text{cos}k_y]$.

The spectrum is given by the absolute value of the eigenvalues of the dynamic matrix $K = (\sigma_3 \otimes \mathbb{I}_2) \mathcal{H}_k$.\cite{Colpa:1978ku}. The Hamiltonian, written in terms of the appropriate Bogoliubov operators $\gamma_k$, is then
\begin{equation}
H =  E_g + \sum_{k, n} \omega_{k,n} \gamma^\dagger_{k,n} \gamma_{k,n},
\end{equation}
where $n$ is the band index and the ground state energy $E_g$ is
\begin{equation}
E_g = E_0 ''-N x A_\theta + \sum_{k,n} \frac{\omega_{k,n}}{2}.
\end{equation}
The canting angle is now determined by minimizing $E_g$ with respect to $\theta$. The effects of the linear boson terms can be seen by comparing the critical value of $W_0^*$ obtained by $E_0''$ and by $E_0'' - N x A_\theta$ in the absence of quantum corrections.  Upon expanding $E_0''$ to leading order in $\theta$ we yield (setting $S$ = 1/2)
\begin{equation}
E_0'' (\theta)/N \approx -\frac{3 J}{2}-B+\left[\frac{B}{2}+\frac{3 J}{2}-\frac{3 W_0}{2}\right] \theta^2,
\end{equation}
whereas the expansion of $E_0''/N - x A_\theta$ is
\begin{equation}
E_0'' (\theta)/N -x A_\theta \approx -\frac{3 J}{2}-B+\left[\frac{B}{4}+J-W_0\right] \theta^2.
\end{equation}
The critical value of $W_0^*$ can be obtained by finding its value when the $\theta^2$ coefficient changes sign.  Hence, without linear boson effects $W_0^* = J + B/3 $, whereas including the linear boson effects reduces the critical value to $W_0^* = J + B/4$.  This technique can be applied to general noncollinear spin systems to identify shifts in critical values of the parameters in the theory.

The dispersion for modest values of $W_0$ consists of two two-fold degenerate branches with a characteristic $d_{x^2-y^2}$ gap--this is due to the translation symmetry breaking nature of the DMI.   For higher period incommensurate triplet density wave states the number of magnon branches will be equal to the period of incommensurability because it is precicely that period which determines the number of sites contained per unit cell. For $\theta = 0$ and $S$ = 1/2 the magnon dispersion is
\begin{equation}
\begin{gathered}
\omega_{k, n} = \frac{1}{2}B + J \pm \\
\frac{1}{2}\sqrt{J^2(\text{cos}k_x+\text{cos}k_y)^2+W_0^2(\text{cos}k_x-\text{cos}k_y)^2}
\end{gathered}
\end{equation}
where the $\pm$ corresponds to the upper and lower band respectively.   
The dispersion is plotted along the high symmetry directions for some representative values of $B$, $W_0$, $\theta$  in Figs. 5-8.
\begin{center}
		 \includegraphics[width=0.45\textwidth]{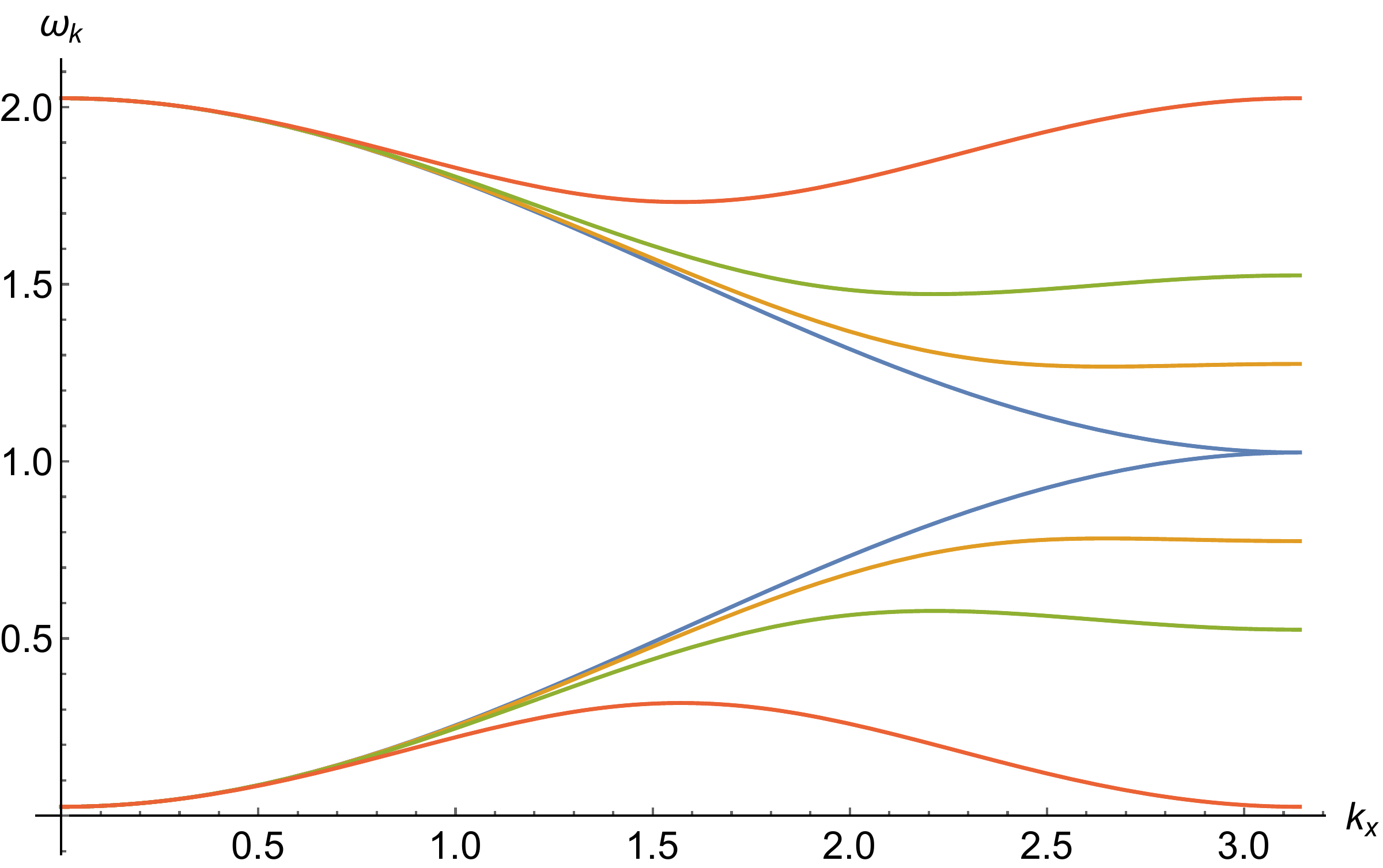}
        \captionof{figure}{
                Magnon dispersion $\omega_{k}$ with $k_y = 0$ in units of $J$ for various values of density wave strength with $B = 0.05J$.  The blue, orange, green, and red curves correspond to $W_0 \rightarrow 0$, $W_0$ = $0.25J$, and $W_0$ = $0.5J$, $W_0$ = $J$ respectively, all below $W_0^*$. As $W_0$ is increased the gap at $k = (\pi, 0)$ increases as 2$W_0$.
        }
        \label{fig:disp1}
\end{center}
\begin{center}
		 \includegraphics[width=0.45\textwidth]{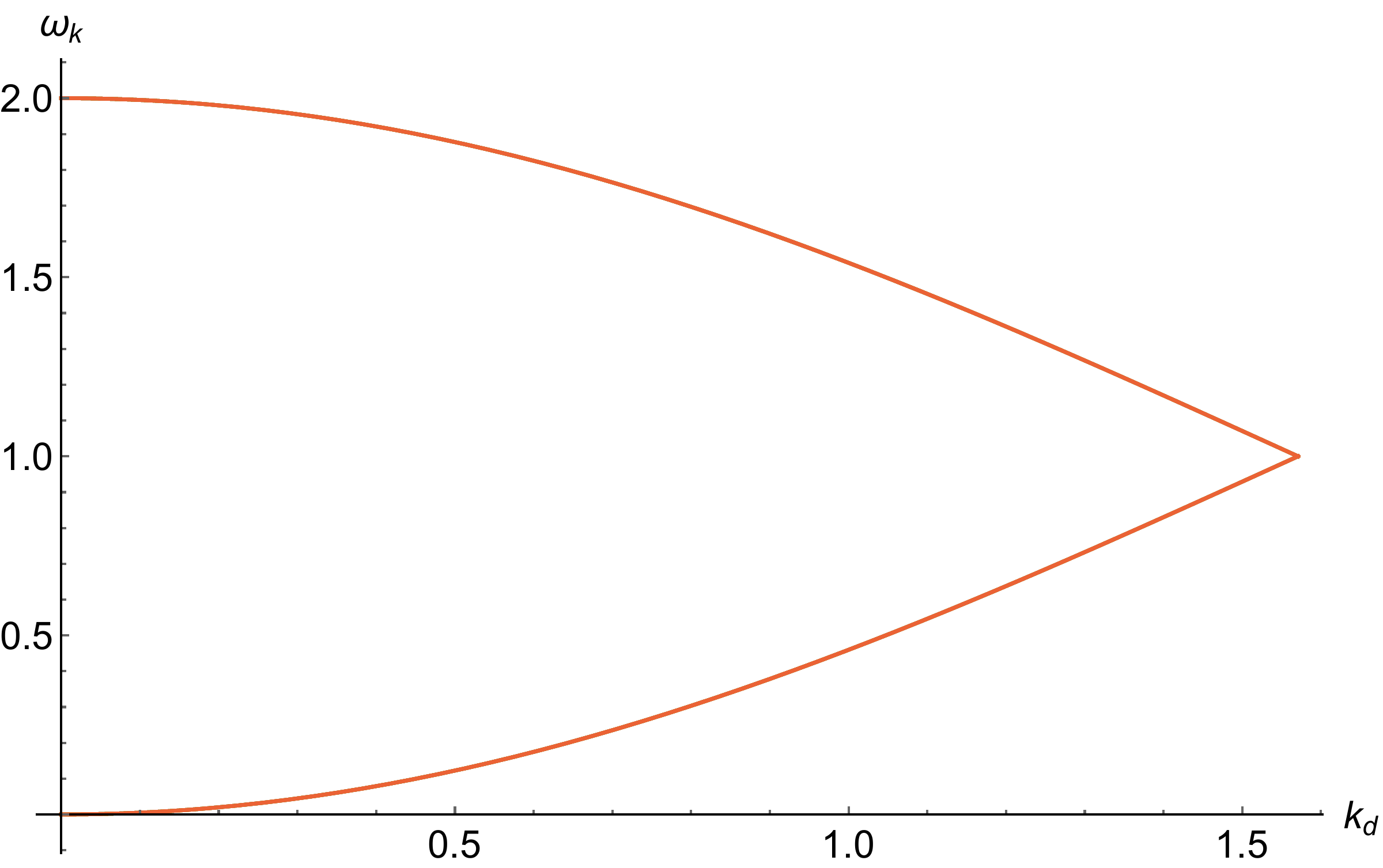}
        \captionof{figure}{
                Magnon dispersion $\omega_{k}$ along the line $k_x = k_y = k_d$ in units of $J$ for various values of density wave strength with $B = 0.05J$.  Tuning $W_0$ does not alter the dispersion in this direction.
        }
        \label{fig:disp2}
\end{center}  
\begin{center}
		 \includegraphics[width=0.45\textwidth]{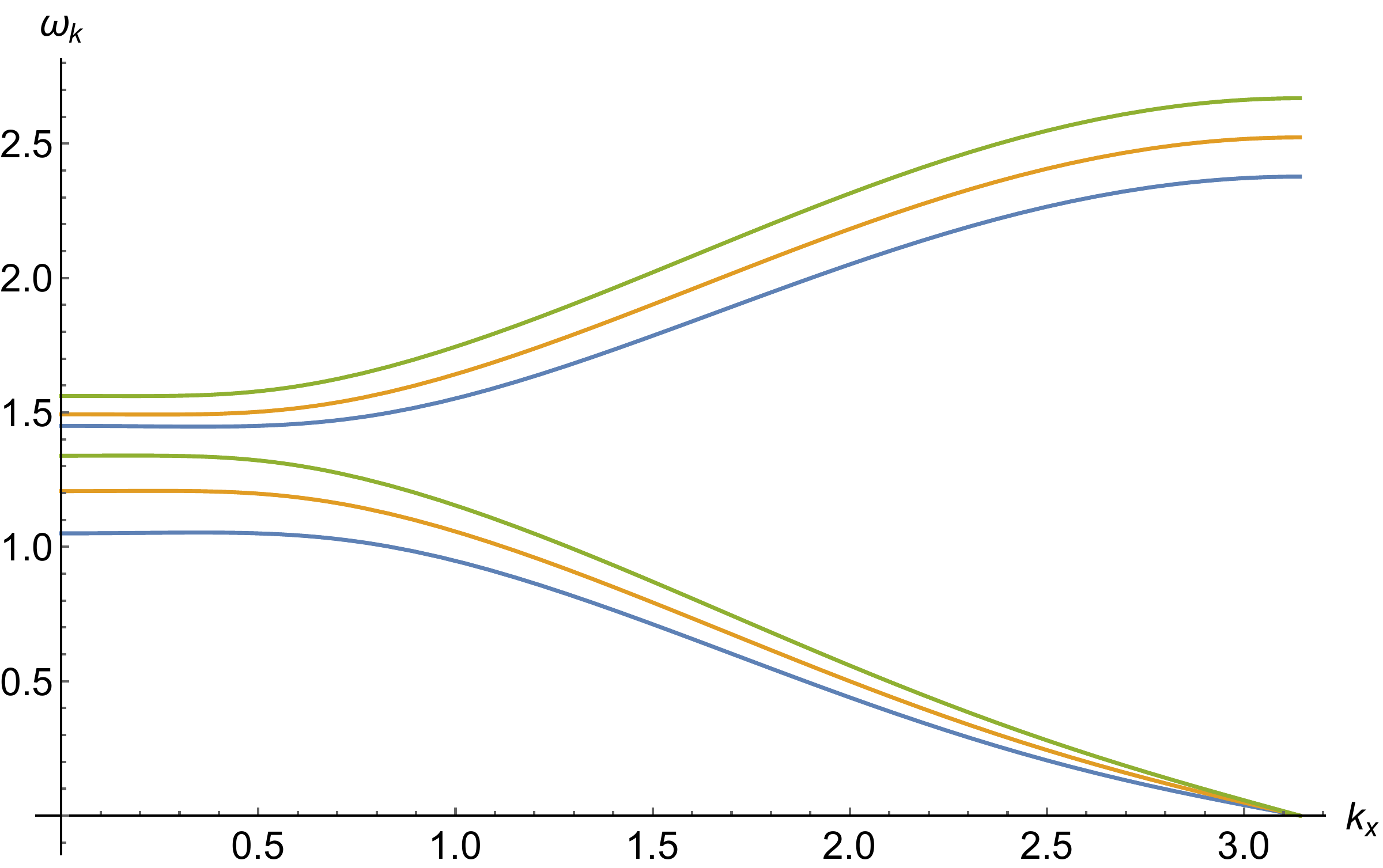}
        \captionof{figure}{
                Magnon dispersion $\omega_{k}$ with $k_y = 0$ in units of $J$ for various values of density wave strength with $B = 0.1J$ with the appropriate canting angles determined by minimization of Eq. 39.  The blue, orange, and green curves correspond to $W_0 \rightarrow1.25J$, $W_0$ = $1.35J$, and $W_0$ = $1.45J$ respectively, all above $W_0^*$. As $W_0$ is increased the gap at $k = (\pi, 0)$ increases as 2$W_0$. The low energy excitations are now governed solely by the points $k = (0, \pi)$ and $(\pi, 0)$. 
        }
        \label{fig:disp3}
\end{center}
\begin{center}
		 \includegraphics[width=0.45\textwidth]{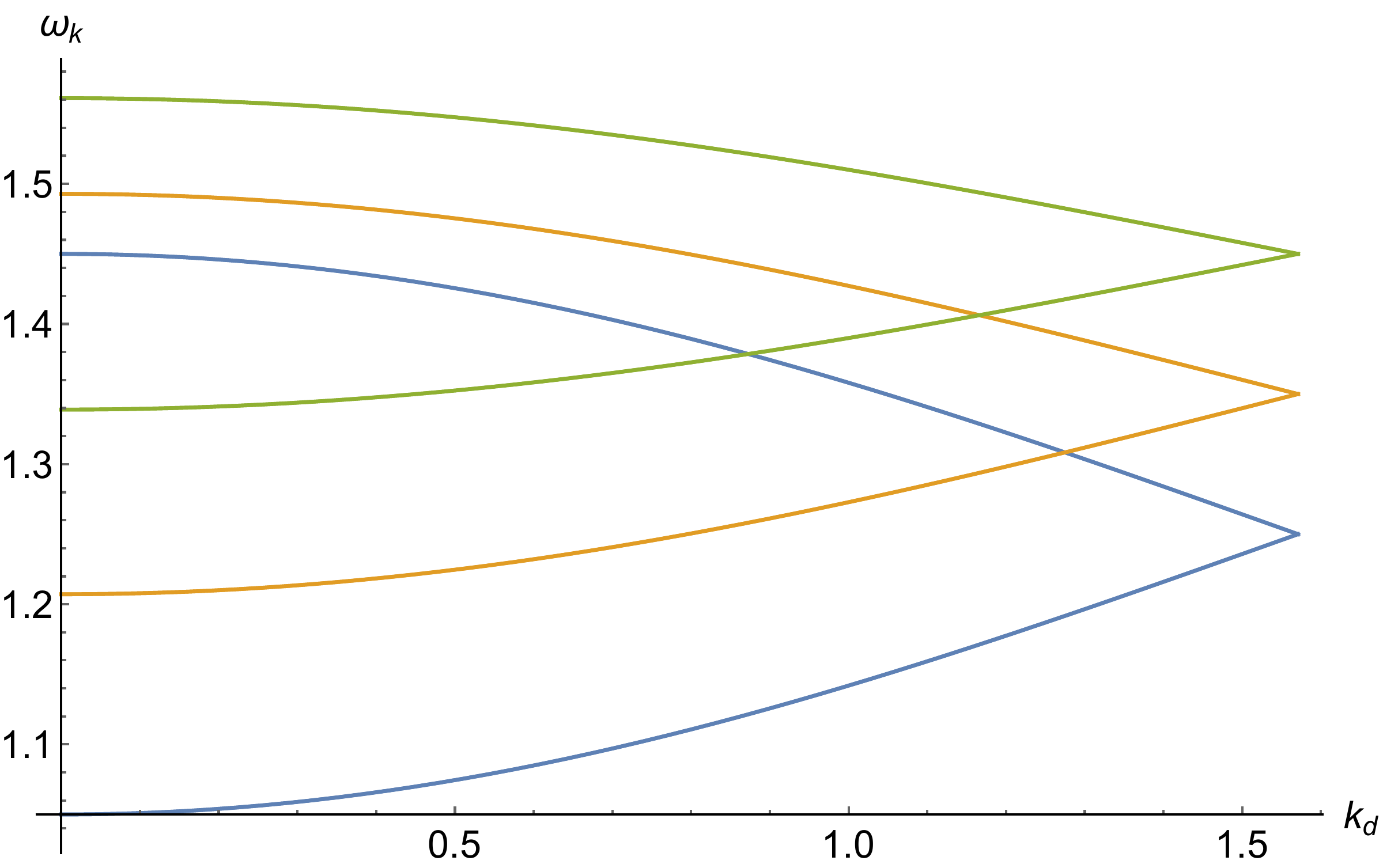}
        \captionof{figure}{
                \label{fig:samplesetup}
                Magnon dispersion $\omega_{k}$ along the line $k_x = k_y = k_d$ in units of $J$ with $B = 0.1J$ with the appropriate canting angles determined by minimization of Eq. 39.  The blue, orange, and green curves correspond to $W_0 \rightarrow1.25J$, $W_0$ = $1.35J$, and $W_0$ = $1.45J$ respectively, all above $W_0^*$. Nonzero canting shifts the spectrum upwards in energy.
        }
        \label{fig:disp4}
\end{center}  
 As $W_0$ is tuned from zero the low energy physics is govened by the $k = (0,0)$ point and gaps develop at $k = (\pi, 0)$, $k = (0, \pi)$ with an energy difference of 2$W_0$.  As $W_0$ approaches its critical value the lowest energy excitations are goverend by $k = (0,0)$, $k = (\pi, 0)$, $k = (0, \pi)$.  Tuning $W_0$ beyond the critical value of density wave strength the low energy excitations are described entirely by the points $k = (\pi, 0)$, $k = (0, \pi)$, and the spectrum is shifted upwards in energy due to the canting of the localized moments.

\section{Discussion}
We have shown that a triplet density wave state induces a DM interaction in the host spin system. The density wave-mediated DM vector is stabilized in topological systems by the direction of the magnetic field, and the symmetry such a DMI is governed entirely by the angular momentum channel of the triplet density wave.  Although it has been shown that the triplet-singlet density wave state produces a nonzero thermal Hall effect\cite{Li:2019wh}, the magnitude of the experimentally-measured thermal Hall effect exceeds the maximum possible contribution from the density wave state alone by an order of magnitude. The excitations of a spin system including DM interactions can, in principle, contribute to the thermal Hall conductivity\cite{Samajdar:2019gk,Han:2019if,Kawano:2019bk}; however, we have shown the particular form of DM interaction generated by the triplet density wave does not seem to produce a nonzero $\kappa_{xy}$, and thus no additional contribution can be found through the influence of the density wave state on the underlying spin system.

Triplet-singlet density wave order is notoriously difficult to detect directly\cite{Hsu:2011ki}, and so it is important to explore possible influences that the state might have on its host system. Experimental detection of such features could, for example, help to assess the importance of the triplet-singlet DDW state in the description of the pseudogap phase of the cuprates. The magnetic structure of LSCO at low doping is Neel order with a small ferromagnetic moment. We have shown that in such a system, the presence of the DDW induces anisotropy in the spin-wave dispersion, reflecting the anisotropy of the DDW.  Furthermore, the magnon branch for such a system has a non-Abelian berry curvature that vanishes upon integration in such a way that $\kappa_{xy}$ = 0.

Additionally, a two patch RG analysis of the $U$-$V$-$J$ model indicates that triplet $d_{x^2-y^2}$-density wave order is can be energetically favorable in a finite region of coupling space given $J/U <0$.\cite{Kampf:2003kx}  Ferromagnetic ordering was also predicted\cite{Kopp:2007fn} to emerge in the highly overdoped cuprates and experimentally confirmed\cite{Sarkar:2020ce} to exist in the CuO$_2$ planes of the cuprates.  We find that the $i \sigma d_{x^2-y^2}+ d_{xy}$-density wave-induced DM interaction in a 2D ferromagnetic system generically produces a magnon spectrum with two branches with a characteristic $d_{x^2-y^2}$ gap.  For higher period incommensurate triplet density wave states in such a spin system the number of magnon branches is equal to the density wave's period of incommensurability.

We have also found that the inclusion of terms in the Hamiltonian linear in Holstein-Primakoff boson operators has a nontrivial effect on the critical behavior of the Hamiltonian. These terms are typically ignored in the literature, which is justified when considering models that are far from the critical regime; however, we have shown that they induce shifts in the critical parameter values which control collinear to noncollinear phase transitions.

\section*{Acknowledgment}
This work was supported in part by funds from David
S. Saxon Presidential Term Chair at UCLA.

\appendix*
\section{Calculating the thermal Hall coefficient}
In our calculation we compute $\kappa_{xy}$ using the Mott-like formula\cite{Wang:2009iv,Qin:2011ea,Qin:2012ck,Matsumoto:2011de,Matsumoto:2014ex}
\begin{equation}\label{eq:thth}
\frac{\kappa_{xy}}{T} = \frac{1}{T^2}\int \frac{(\epsilon-\mu)^2}{\text{cosh}^2\left(\frac{\epsilon-\mu}{2T}\right)} \sigma_{xy} (\epsilon) d\epsilon
\end{equation}
\vspace{.3mm}
where $\mu$ is the chemical potential and $\sigma_{xy}(\epsilon)$ is the Hall coefficient for the system at zero temperature with chemical potential $\epsilon$.  We implement the linear-in-field approximation where calculation of $\kappa_{xy}$ is greatly simplified at low magnetic field:\cite{Yang:2020gi}
\begin{equation}
\sigma_{xy}(\epsilon) \approx B \partial_{B} \sigma_{xy}(\epsilon) |_{B=0} = - B \tilde{\mathcal{B}}(\epsilon)
\end{equation}
where $B$ is magnitude of the magnetic field, and $\tilde{\mathcal{B}}(\epsilon)$ is the effective Berry curvature density given by
\begin{equation}
\tilde{\mathcal{B}}(\epsilon) = \sum_{n k s} \mathcal{B}_{n s k} \delta(\epsilon - \epsilon_{n s k}).
\end{equation} 
Here $n = \pm 1$ for the lower and upper bands, and the spin index $s = \pm 1$.
The Berry curvature and dispersion for the mixed triplet-singlet DDW are given by\cite{Hsu:2011ki}
\begin{equation}
\mathcal{B}_{n s k} = n s \frac{t W_0 \Delta_0}{E_k^3}(\sin^2 k_y + \sin^2 k_x \cos^2 k_y),
\end{equation}
\begin{equation}
\epsilon_{n s k} = \epsilon_{2 k} - n E_k,
\end{equation}
 with $E_k$ is defined as 
\begin{equation}
E_k = \sqrt{4t^2(\cos k_x+\cos k_y)^2 + W_k^2 + \Delta_k^2}.
\end{equation}
where 
\begin{equation}
\begin{split}
W_k = \frac{W_0}{2}(\cos k_x - \cos k_y)\\
\Delta_k = \Delta_0 \sin k_x \sin k_y,
\end{split}
\end{equation}
and 
\begin{equation}
\epsilon_{2k} = 4 t' \cos k_x \cos k_y.
\end{equation}
Upon integration over $\epsilon$, Eq. \eqref{eq:thth} simplifies to
\begin{equation}
\frac{\kappa_{xy}}{T} = \frac{B}{2 T^3}\sum_{\substack{k \in RBZ \\ \alpha = \pm}} \mathcal{B}_{+ + k}\left[\frac{-\alpha(\epsilon_{\alpha \alpha k}-\mu)^2}{\cosh^2(\beta (\epsilon_{\alpha \alpha k}-\mu)/2 )}\right]
\end{equation}
which may be evaluated with ease.  We numerically integrate this quantity and plot it in Fig. \ref{fig:kappaXY}

\bibliography{DDW_DMI_star}

\end{document}